\documentstyle[12pt]{article}

\begin{document}

\begin{center}
{\bf CLASSICAL STOCHASTIC APPROACH TO COSMOLOGY REVISITED }

\vspace{1cm} Moncy V John{\bf $^{\dag }$ Sivakumar C$^{\ddag }$ and K. Babu
Joseph}\\ Department of Physics, Cochin University of Science and
Technology, Kochi 682022, India

\vspace{1cm} {\bf Short title :} Stochastic Approach To Cosmology

\vspace{1cm} {\bf Abstract}
\end{center}

The classical stochastic model of cosmology recently developed by us is
reconsidered. In that approach the parameter $w$ defined by the equation of
state $p=w{\rho }$ was taken to be fluctuating with mean zero and we
compared the theoretical probability distribution function(PDF) for the
Hubble parameter with observational data corresponding to a universe with
matter and vacuum energy. Eventhough qualitative agreement between the two
was obtained, an attempt is herein made to introduce a more realistic
assumption for the mean of $w$ and use it for the calculations. In the
present theory the mean values of both $p$ and $w$ are taken to be nonzero.
The theoretical and observational PDFs are compared for different epochs and
values of the Hubble parameter. The corresponding values of the diffusion
constant $D$ obtained are approximately constant. We use the scatter in the
observed redshift-magnitude data of Type Ia supernova to place limits on the
stochastic variation in expansion rate and consequently, on the stochastic
variation of the equation of state.

\vspace{1cm} \noindent {PACS numbers:} 98.65 Dx, 98.80 Es, 02.50 Ey

\vspace{0.5cm} \noindent 
\noindent {\dag} Permanent Address: Department of Physics, St. Thomas
College, Kozhencherri - 689641, Kerala, India. Email : moncy@stthom.ernet.in

\noindent {\ddag } e-mail: sivakumarc@cusat.ac.in \newpage
\baselineskip24pt

\section{Introduction}

The uncertainty in the determination of the Hubble parameter H, which is a
measure of the expansion rate of the universe, is one of the most intriguing
issues in the history of cosmology. The origin of the uncertainty is obvious
from the redshift-magnitude diagram (Hubble diagram); despite rigorous
attempts to control the random errors in measurement, there is a clear
scatter in it, though it is now possible to narrow down this to a great
extent. But, as we show later in this paper by using the redshift-apparent
magnitude data for Type Ia supernovae (SNe Ia) \cite{perl99}, the scatter
increases as we go to higher redshifts. In a recent work \cite{siva}, we
have attempted to explain this scatter as arising from an inherent
stochastic or nondeterministic nature of the Hubble parameter. It was shown
that a fluctuating $w$-factor in the equation of state $p=w\rho $ will lead
to this kind of behaviour for $H$. In order to arrive at the notion of a
stochastic equation of state, we consider the following. Several recent
measurements \cite{perl99,riess98,freedman,kamion,clad} indicate that the
present universe contains components other than ordinary matter and
radiation, like vacuum energy, quintessence etc.. Let the equation of state
for each component, with density $\rho _i$, be written as $p_i=w_i\rho _i$, (%
$i=1,2,..$), where $p_i$ is the corresponding pressure ($w_i=0$ for dust, $%
1/3$ for radiation, $-1$ for vacuum energy, etc.; in general, $-1<w_i<+1$).
The total energy density is $\rho =\sum {_i}\rho _i$ and the total pressure $%
p=\sum_ip_i$. From the law of conservation of total energy-momentum tensor
expressed in the form 
\begin{eqnarray}
\dot{\rho }=\sum _i \dot{\rho }_i = -3\frac {\dot{a}}{a}(\rho +p) &
=&-3\frac{\dot{a}}{a}\rho (1+w) \\ \nonumber
&=&-3\frac{\dot{a}}{a}\sum _i \rho_i (1+w_i ),
\end{eqnarray}

\noindent which follows from the Einstein equation and one can write the
total $w$-factor as 
\begin{equation}
\label{eq:w}w\equiv \frac p\rho =-\frac{\sum_i\dot {\rho _i}}{3(\dot
a/a)\rho }-1=\sum_i\frac{\rho _i}\rho (1+w_i)-1, 
\end{equation}
\noindent where $a$ is the scale factor. The conservation of individual
components, which may be expressed as $\dot \rho _i=-3(\dot a/a)\rho
_i(1+w_i)$, is only an extra assumption since it does not follow from the
Einstein's field equation. Equivalently, it can be stated that in a many
component fluid as in the above case, the Einstein equations, along with the
equations of state of individual components are insufficient to determine
the individual $\dot {\rho _i}$'s. Thus it is more general not to assume
conservation of individual components and this will lead to creation of one
component at the expense of other components. Since they are not uniquely
determined by the field equations, such creation can be considered as
sporadic events, like those occurring in galactic nuclei, which can result
in fluctuations in the ratio $\rho _i/\rho $. Here we make the assumption
that this will lead to a stochastic equation of state, as seen from equation
(\ref{eq:w}) above. Consequently, also the expansion rate will be
fluctuating and the equation for the Hubble parameter will appear as a
Langevin type \cite{risken} equation. Recently Padmanabhan \cite{paddy} has
proposed a similar scenario in which the observed cosmological constant
arises from quantum fluctuations of some energy density; consequently the
FRW equations are to be solved with a stochastic term on the right hand
side. In \cite{siva}, it was assumed for the sake of simplicity that space
sections are flat and $w$ is a Gaussian $\delta $-correlated stochastic
force with zero mean. With these assumptions, we have written the
Fokker-Planck (FP) equation, whose solution gives the theoretical
probability distribution function (PDF) for $H_0$ at time $t_0$, denoted as $%
W(h,t_0)$ (where $H_0=100h$ km s$^{-1}$ Mpc$^{-1}$; the subscript 0 denotes
the present epoch). Using the redshift-apparent magnitude data $\mu _o$ for
SN Ia used in \cite{perl99}, we computed the observational PDF $p(h|\mu _o)$
for $h$ in the present universe, again assuming its space sections to be
flat. This PDF arises from the point-to-point variance of the Hubble flow.
We compared the two plots for the present universe and found them to agree
well, for a value of the diffusion constant $D$ appearing in the FP equation
equal to $4\times 10^{13}$ s.

This result is a first step towards an understanding of the anomalous
scatter in the Hubble diagram at high redshifts. However, there are certain
refinements to be made in our analysis. One drawback of the above scheme of
comparing these two PDF's is that when we derived $W(h,t)$, the assumption
was made that $w$ has mean value zero, whereas the observational PDF $%
p(h|\mu _o)$ was evaluated for a model which contains dust and vacuum
energy, which has the mean total pressure negative. Instead, if we had used
in this evaluation the expression for the distance modulus for a flat
universe which is dust dominated (i.e., with $w=0$), an observational PDF
will be obtained, but the best fit value for $h$ would be ridiculously low.
(This is all the fuss about the new observations - that they are
incompatible with an $\Omega _\Lambda =0$ flat model.)

Another shortcoming is that though in both cases we take the PDF for $h$, it
remains to be explained how legitimate is the comparison of $W(h,t_0)$ for
the present universe with a PDF $p(h|\mu_0)$ evaluated using the data that
include high redshift objects, which belong to the distant past.

In this paper, we attempt to rectify these two defects and to make a more
rigorous test of the stochastic assumptions using observational data by (1)
comparing both the theoretical and observational PDF's evaluated for the
same model, which is an alternative flat model \cite{mvj} and (2) evaluating
the observational PDF $p(h_j |\mu_{oj})$ for the Hubble parameter at the
same epoch $t_j$ as that in the theoretical PDF $W(h_j ,t_j)$. This
procedure helps us to compare the theoretical and observational PDF's for
the Hubble parameter for the same model, and at the same epoch. The value of
the diffusion constant evaluated at any time is obtained as nearly a
constant, in agreement with our assumptions. A novel feature in our present
approach is that we evaluate the observational PDF for Hubble parameter at
various instants in the past, also with an objective of justifying our
assertion that the scatter increases as we go into the past.

The paper is organised as follows. Section 2 gives a brief review of the
alternative model to be used and then develops the stochastic approach in
it. Section 3 gives the new technique of finding the PDF for $h$ at any time
in the past or present epochs and the results obtained on comparison between
the theoretical and observational PDF's. The conclusions derived from it are
discussed in section 4.

\section{Stochastic approach in the new model}

In all FRW models, the Einstein equations, when combined with the
conservation of total energy-momentum tensor can be written in terms of the
Hubble parameter as 
\begin{equation}
\dot H=-H^2-\frac{4\pi G}{3} \left( \rho+3p\right) . 
\end{equation}
Overdot denotes time derivative. If we restrict ourselves to flat models,
then (with $p=w\rho$), 
\begin{equation}
\label{eq:hdot}\dot H=-\frac 32H^2\left( 1+w\right) . 
\end{equation}

In \cite{siva}, we considered this flat case and assumed that $w$ is a
Gaussian $\delta $-correlated Langevin force term with mean value zero. This
means that the mean total pressure of the universe is zero, the same as that
for dust. But many recent observations are incompatible with this model and
hence, as mentioned in the introduction, we look for a more observationally
correct, but simple model to apply our stochastic treatment.

The deterministic model \cite{mvj} we propose to use is the one in which the
total energy density obeys the condition $\rho +3p=0$ and hence having a
coasting ($a\propto t$) evolution. This model is derived on the basis of
some dimensional considerations in line with quantum cosmology. If we assume
that the energy components in this model are dust and vacuum, then the above
condition gives $\rho _m/\rho _v=2$ and if they are only radiation and
vacuum, then $\rho _r/\rho _v=1$. In \cite{mvj}, it was shown that in this
model, most outstanding cosmological problems like those of flatness,
horizon, monopole, entropy, size and age of the universe, and the
cosmological constant are absent. It was also shown that this model can
solve the problem of generation of density perturbations at scales well
above the present Hubble radius and that it can generate such density
perturbations even after the era of nucleosynthesis. Though it is mentioned
in the paper that recent observations on the redshift-apparent magnitude
relation are at variance with the predictions of this model, in a recent
communication \cite{mvj1}, this issue was studied in detail and found that
those observations do not provide any strong evidence against it.

It may be noted that the underlying model described above deviates from
`main stream' cosmology to some extent and the most significant cosmological
observations such as nucleosynthesis, large scale structures and microwave
background radiation, in this model are not well-studied. However, to repeat
our statements above, we choose this model to implement our stochastic
approach mainly due to its simplicity and its capability to explain the
recent supernova data.

In view of the fact that this model has $w=-1/3$ in the deterministic case,
we rewrite the above equation (\ref{eq:hdot}) with $w^{\prime }\equiv
w+(1/3) $, as 
\begin{equation}
\label{eq:hdot1}\dot H=-\frac 32H^2\left( \frac 23+w^{\prime }\right) . 
\end{equation}
We now assume that $w^{\prime }$ fluctuates about its zero mean value and is 
$\delta $-correlated. Making a substitution 
$$
x\equiv \frac 1H 
$$
the above equation becomes 
\begin{equation}
\label{eq:xdot}\dot x=1+\frac 32w^{\prime }. 
\end{equation}
When $w^{\prime }=0$, this equation is analogous to that of a particle
moving in a medium with constant velocity. With a fluctuating $w{\prime }$,
the analogous particle is subjected to random forces as it moves. We write
the FP equation for the distribution function for the variable $x$ as \cite
{risken} 
\begin{equation}
\label{eq:FPE}\frac{\partial W^{\prime }}{\partial t}(x,t)=\left[
-D^{(1)}\frac \partial {\partial x}+D^{(2)}\frac{\partial ^2}{\partial x^2}%
\right] W^{\prime }\left( x,t\right) , 
\end{equation}
the solution of which gives the PDF $W^{^{\prime }}$ for the variable $x$ at
time $t$. Here the drift coefficient $D^{(1)}$ is the constant velocity
term, equal to unity, and the diffusion coefficient $D^{(2)}\equiv D$ is
assumed to have some constant value, to be determined from observation. The
FP equation is solved by first assuming that the variables are separable; 
\begin{equation}
\label{eq:W}W^{\prime }\left( x,t\right) =\phi _n\left( x\right) e^{-\lambda
_nt}, 
\end{equation}
where $\phi _n\left( x\right) $ and $\lambda _n$ are the eigen functions and
eigen values of the Fokker-Planck operator 
\begin{equation}
L_{FP}=\left[ -\frac{\partial D^{(1)}}{\partial x}(x)+\frac{\partial
^2D^{(2)}}{\partial x^2}(x)\right] . 
\end{equation}
Now we define two more functions in order to get a solution for the FP
equation; 
\begin{equation}
\Phi \left( x\right) \equiv -\int \frac{D^{(1)}}{D^{(2)}}dx^{^{\prime
}}=-\frac xD 
\end{equation}
and 
\begin{equation}
\label{eq:psi}\psi _n(x)=\exp \left( \frac \Phi 2\right) \phi _n(x)=\exp
\left( -\frac x{2D}\right) \phi _n(x), 
\end{equation}
where $\Phi \left( x\right) $ is treated as a stochastic potential and $\psi
_n\left( x\right) $ are the eigen functions of the Hermitian operator $L$
defined as 
\begin{equation}
L=\exp \left( \frac \Phi 2\right) L_{FP}\exp \left( -\frac \Phi 2\right) . 
\end{equation}
Making use of equations (\ref{eq:W}) and (\ref{eq:psi}), the time
independent part of FP equation becomes 
\begin{equation}
\frac{\partial ^2\psi _n}{\partial x^2}(x)=\left[ \frac 1{4D^2}-\frac{%
\lambda _n}D\right] \psi _n\left( x\right) =-k^2\psi _n\left( x\right) . 
\end{equation}
Here, 
\begin{equation}
k=\pm \left[ \frac{\lambda _n}D-\frac 1{4D^2}\right] ^{\frac 12}. 
\end{equation}
The most general solution to (\ref{eq:FPE}) is 
\begin{equation}
W^{\prime }\left( x,t\right) =\sum_{n=0^{}}^\infty c_ne^{-\lambda _nt}\phi
_n\left( x\right) . 
\end{equation}
For $\lambda _n<\frac 1{4D}$, the solution $\psi _n\left( x\right) $ is
exponentially diverging, which is not an admissible solution. Thus we
conclude that $\lambda _n\geq \frac 1{4D}$ so that 
\begin{equation}
\phi _k(x)=A\exp \left( \frac x{2D}+ikx\right) . 
\end{equation}
We make use of the completeness relation 
\begin{equation}
\delta (x-x^{\prime })=\int_{-\infty }^{+\infty }\psi _k^{*}(x)\psi
_k(x^{\prime })dk 
\end{equation}
to specify the initial condition $x=x^{\prime }$ at $t=t^{\prime }$. The
transition probability for the variable to change from $x^{\prime }$ at time 
$t^{\prime }$ to $x$ at time $t$ is \cite{risken} 
\begin{eqnarray}
P( x,t\mid x^{\prime
},t^{\prime }) & = & e^{L_{FP}(t-t^{\prime })}\delta (
x-x^{\prime }) \\  \nonumber & = & \exp \left[ \frac{\Phi (x^{\prime
})}2-\frac{\Phi (x)}2\right] \int_{-\infty }^{+\infty } dk \psi
_k^{*}(x) \psi _k ( x^{\prime }) 
\exp \left[ -\lambda (k)( t-t^{\prime }) \right] \\  \nonumber & = &
\frac 1{2\sqrt{\pi D ( t-t^{\prime }) }}\exp \left[
-\frac{\left[ (x-x^{^{\prime }})-( t-t^{\prime })
\right] }{4D ( t-t^{\prime }) }^2\right] ,  \label{eq:P}
 \end{eqnarray}  
where we have chosen $A=\frac 1{\sqrt{2\pi }}$ for normalisation purpose.
For the special initial condition $W^{\prime }(x,t^{\prime })=\delta
(x-x^{\prime })$, the transition probability $P(x,t\mid x^{\prime
},t^{\prime })$ is the distribution function $W^{\prime }(x,t)$ \cite{risken}%
. In our case, we have the initial condition $x=x^{\prime }=0$ at $%
t=t^{\prime }=0$, so that 
\begin{equation}
\label{eq:Wxt}W^{\prime }\left( x,t\right) =P(x,t|0,0)=\frac 1{2\sqrt{\pi Dt}%
}\exp \left[ -\frac{\left( x-t\right) }{4Dt}^2\right] , 
\end{equation}
which is in Gaussian form and has its peak moving along in such a way that
the expectation value of the variable is $\langle x\rangle =t$. This
corresponds to the deterministic solution of (\ref{eq:xdot}). The width of
the Gaussian is found from the variance $\sigma ^2=\left\langle \left(
x-\left\langle x\right\rangle \right) ^2\right\rangle =2Dt$ and we find $%
\sigma \geq \langle x\rangle $ till $t=2D$. We can immediately rewrite this
distribution function in terms of the stochastic Hubble parameter $H$ as 
\begin{equation}
\label{eq:WHt}W^{\prime \prime }(H,t)=\frac 1{2H^2}\frac 1{\sqrt{\pi Dt}%
}\exp \left[ -\frac{\left( 1-Ht\right) ^2}{4H^2Dt}\right] . 
\end{equation}
With $H=100\;h$ km s$^{-1}$ Mpc$^{-1}$, $t=t_{17}\times 10^{17}$ s and $%
D=D_{17}\times 10^{17}$ s, the PDF $W(h,t)$ can be written as 
\begin{equation}
\label{eq:Wht}W(h,t)=\frac{3.0856}{2h^2}\frac 1{\sqrt{\pi D_{17}t_{17}}}\exp
\left[ -\frac{\left( 3.0856-ht_{17}\right) ^2}{4h^2D_{17}t_{17}}\right] . 
\end{equation}
For the range of values of interest, $1<t_{17}<5$ and $D_{17}\sim 10^{-3}$, $%
W(h,t)$ is approximately a Gaussian. For fixed $D$, the half width of the
Gaussian is found to increase as we go to lower values of $t$.

\section{PDF for $H$ from observational data}

From the Hubble diagram, one can find the PDF for the present Hubble
constant $H_0$ in the following way. The traditional measure of distance to
a SN is its observed distance modulus $\mu_o = m_{bol} -M_{bol}$, the
difference between its bolometric apparent magnitude and absolute magnitude.
In the FRW cosmology, the distance modulus is predicted from the source's
redshift, $z$, according to 
\begin{equation}
\label{eq:mup}\mu_p = 5 \log \left[ \frac{D_L}{1 \hbox{Mpc}}\right] +25, 
\end{equation}
where $D_L$ is the luminosity distance, found as 
\begin{equation}
\label{eq:dl1}D_L = r_j a(t_0 )(1+z) . 
\end{equation}
$a(t_0 )$ is the present scale factor and $r_j$ is the radial coordinate of
the SN Ia which emitted the light at some time $t_j$ in the past. In flat
FRW models, $r_j$ is found as 
\begin{equation}
r_j = \int _{t_j }^{t_0 }\frac{c\; dt}{a(t)}, 
\end{equation}
For the coasting model discussed in the previous section, for $k=0$, $r_j$
can be evaluated as 
\begin{equation}
\label{eq:r1}r_j = \frac{c t_0}{a(t_0 )}\int _{t_j }^{t_0 }\frac{ dt}{t}
=\frac {ct_0 }{a(t_0 )} \ln (1+z) , 
\end{equation}
so that 
\begin{equation}
\label{eq:dl2}D_L = \frac {c(1+z)}{H_0 }\ln (1+z). 
\end{equation}
One can use this expression in (\ref{eq:mup}) to obtain the predicted
distance modulus of an object with redshift $z$. Conventionally, assuming
that the observed and predicted distance moduli coincide, one can find a
value of $H_0$. For a collection of objects, also one can find the
likelihood for $H_0$, from a $\chi ^{2}$ statistic; 
\begin{equation}
\chi ^{2} = \sum_i \frac {(\mu_{p,i} -\mu_{o,i})^2}{\sigma _i^2 }, 
\end{equation}
where $\sigma _i $ is the total uncertainty in the corrected peak magnitude
of SN Ia. For the special model we are considering, $h$ is the only
parameter and the normalised PDF can now be obtained as \cite{riess98} 
\begin{equation}
\label{eq:phmu}p(h|\mu_0) = \frac {\exp (-\chi ^2 /2)}{\int
_{-\infty}^{\infty} dh \exp -\chi^2 /2}. 
\end{equation}

As in \cite{siva}, we compute $p(h|\mu_o)$ for the new model using the SNe
data in \cite{perl99}, which corresponds to their Fit C and attempt to
compare $p(h|\mu_o)$ with the PDF $W(h , t_0 )$ to evaluate the diffusion
constant $D$ appearing in this expression. It is found that the two curves,
shown in figure 1, coincide for a value of $D\approx 2.36\times 10^{13}$ s.
(This corresponds to an age $4.8583 \times 10^{17}$ s.) Since our primary
objective is to make an order of magnitude evaluation of $D$, we chose a
fiducial absolute magnitude for SN Ia in computing $\mu_o$ , equal to $-19.3$
mag. Slight variations in this quantity will not significantly affect $D$,
though the best fit value for $h$ may change.

In the above, we compared the theoretical and observational PDF's for the
same alternative model and thus it does not have the first shortcoming
mentioned in the introduction. The other incompatibility which still exists
can be explicitly stated as follows: $W(h ,t_0 )$ is the PDF for the Hubble
parameter of the present universe and it contains the diffusion constant $D$
. But $p(h |\mu_o )$, which we attempt to identify with $W(h,t_0)$, depends
on the scatter in the Hubble diagram for all ranges of redshift. For
instance, if we include more high redshift objects in our sample, the
scatter would be larger and hence the half-width of the distribution $p(h
|\mu_o )$ will be larger. This, in turn, will affect the computed value of $%
D $, which is quite unreasonable.

This problem can, however, be overcome if we agree to compute $p(h_j|\mu
_{oj})$ for each value of redshift $z$ (or for small enough redshift
intervals centred about such values), and compare these with $W(h_j,t_j)$
that corresponds to the same epoch $t_j$. To do this, we modify (\ref{eq:dl2}%
) by re-evaluating $r_j$ in (\ref{eq:r1}) in a different way. One can also
write, for the new deterministic model 
\begin{equation}
\label{eq:r12}r_j=\frac{ct_j}{a(t_j)}\int_{t_j}^{t_0}\frac{dt}t=\frac
c{H_ja(t_j)}\ln (1+z), 
\end{equation}
so that 
\begin{equation}
\label{eq:dl3}D_L=\frac{c(1+z)^2}{H_j}\ln (1+z). 
\end{equation}
Evaluating $\mu _p$ using this expression in (\ref{eq:mup}), we can
calculate $\chi ^2$ and hence also $p(h_j|\mu _{oj})$, which is the PDF for
the Hubble parameter at some particular value of $z$. We divide the data in 
\cite{perl99} for various ranges around $z$ = 0.05, 0.15, 0.35, 0.45, 0.55
and 0.65, each with $\Delta z$ = 0.05. The PDF for the average Hubble
parameter for such intervals is calculated with an expression identical to ( 
\ref{eq:phmu}). The results are plotted in figure 2 along with the
corresponding $W(h_j,t_j)$ which overlaps with them. The relevant parameters
are given in table 1.

\section{Conclusions}

It can be noted from figure (2) and table 1 that for the intervals with
larger values of $z$, the 68.3\% credible region of $p(h|\mu _{oj})$ has a
half-width $\sigma _h$ which also increases (The intervals with centre at $%
z=0.15$ and 0.35 are exceptions to it, but this may be due to the fact that
these intervals contain only very few objects. As more SN Ia are observed in
these redshift bins, an accurate picture will emerge.). This behaviour is
the one expected from theory, as noted while plotting the theoretical PDF (%
\ref{eq:Wht}). The value of the diffusion constant $D$ evaluated for various
intervals, however, does not show any dependence on $z$. This justifies our
assumption (taken for the sake of simplicity) that $D$ is some constant.

As remarked in the introduction, a novel feature of the present analysis is
that we compute the PDF for $H$ at various epochs in the past. However,
there is a limitation, even in the present analysis; the intervals we
consider are with $\Delta z=0.05$ and this value may not be small enough to
give correct answers. The procedure can be improved very much in the future,
when the number of observed SN Ia becomes large.

It should be kept in mind that the measurements are made with finite
accuracy and hence have a scatter arising from intinsic statistics of error.
Here we have not shown that the scatter in the Hubble diagram is
significantly more than that expected from the known errors. The present
theory can survive only if $p(h\mid \mu _o)$ does not become narrower with
increase in accuracy of measurement. Until it is confirmed by future
observations like the Supernova Acceleration Probe (SNAP) \cite{weller},
which aims to give precise luminosity distance of $\approx 2000$ SN Ia upto $%
z=1.7$ every year, the computed value of the diffusion constant can not be
claimed as accurate. Given this circumstance, it is safer to conclude for
the present that the scatter in the SN\ Ia data places limits on the
stochastic variation in the expansion rate and consequently, on the
stochastic variation of the equation of state. In other words, it is
justifiable to consider the value of $D$ computed by us using SN Ia data as
an upper limit to a possible diffusion constant for the universe.

\vspace{.75cm} \noindent{\bf Acknowledgment}

We are thankful to Professor Jayant Narlikar for valuable discussion and to
the referee for his suggestions. CS thanks CSIR, New Delhi for the award of
a Research Fellowship.

\newpage 

\begin{center}
Table 1
\end{center}

\begin{tabular}{l|l|l|l|l|l} \hline
Redshift & No. of SNe & Best fit & Standard & Age in & Diffusion \\ 
z & in the & value of & deviation & units of 10$^{17}$ s& constant \\ 
     & interval $z\pm 0.05$& h& $\sigma_h $& $t_{17}$ & D s \\
\hline 0.05 & 15 & 0.693 & 0.011 &4.4502 & 0.5775$\times 10^{14} $ \\
0.15 & 3 &0.772 & 0.025 &3.9987 & 2.147  $\times 10^{14} $ \\
0.35 & 5 &0.830 &0.024 &3.7204 & 1.5  $\times 10^{14} $ \\
0.45 &15 &0.875 &0.017 &3.528 &  1.655  $\times 10^{14} $ \\
0.55 &7 &0.985 &0.026 &3.1345 & 1.12  $\times 10^{14} $ \\
0.65 &6 &1.043 &0.030 &2.959 &  1.226  $\times 10^{14} $ \\ \hline
\end{tabular}

\newpage\ 

\noindent Figure caption:

\noindent Figure 1: Observational and theoretical PDF's vs $h$, using the
apparent magnitude-redshift data for Type Ia supernovae as given in [1],
which corresponds to their Fit C. The continuous line is for the
observational PDF whereas the dotted line gives the theoretical PDF.

\vspace{.5cm}

\noindent Figure 2: Observational and theoretical PDF's vs $h _j$ (where $h
_j$ corresponds to the Hubble parameter for the epoch centered about
redshifts $z=0.05, 0.15, 0.35, 0.45, 0.55, 0.65$), using the apparent
magnitude-redshift data for Type Ia supernovae as given in [1], which
corresponds to their Fit C and which lies in the interval $z \pm 0.05$.

\end{document}